\documentclass{webofc}
\usepackage[varg]{txfonts}   
\usepackage{subcaption}
\usepackage{amsmath,amssymb}
\begin{document}
\title{Studying strangeness and baryon production in small systems through $\bf \Xi-$hadron correlations using the ALICE detector}
%
%

\author{\firstname{Jonatan} \lastname{Adolfsson}\inst{1}\fnsep\thanks{\email{jonatan.adolfsson@hep.lu.se}} (for the ALICE Collaboration)}

\institute{Lund University, Lund, Sweden}

\abstract{%
These proceedings summarise recent measurements of angular correlations between the $\Xi$ baryon and identified hadrons in pp collisions at $\sqrt{s}=13\,\mathrm{TeV}$ using the ALICE detector. The results are compared with both string-based (PYTHIA8 with extensions) and core-corona (EPOS-LHC) models, to improve our understanding of strangeness and baryon production in small systems. The results favour baryon production through string junctions over diquark breaking, but the PYTHIA models fail at describing the relatively wide $\Xi-$strangeness jet peak, indicating stronger diffusion of strange quarks in data. On the other hand, EPOS-LHC is missing local conservation of quantum numbers, making it difficult to draw any conclusion about the core-corona model. 
}
\maketitle
%
\section{Introduction}
\label{sec:introduction}
Today, it is widely accepted that a quark--gluon plasma (QGP) is formed in heavy-ion collisions. What is not yet understood, however, is that several QGP key signatures have also been observed in small systems such as high-multiplicity proton--proton (pp) collisions, where the QGP is not expected to form. One such signature is the enhanced yields of multi-strange baryons, which seem to scale smoothly with system size~\cite{ALICE nature paper}.

Several phenomenological models are being developed to try to understand the origin of this. Two different approaches are being explored. In one approach, microscopic models based on colour strings are extended with new mechanisms for strange-baryon formation and other features to mimic collective behaviour. This approach is used in PYTHIA \cite{Sjostrand PYTHIA8 manual}, where baryons in the standard configuration (Monash tune) are formed through diquark breaking. This is extended by allowing string junctions \cite{Skands junctions} and -- as a further extension -- rope hadronisation \cite{Bierlich ropes}, where the latter provides strangeness enhancement. The other approach is the core-corona model, which is based on a two-phase state with a QGP-like core and a string-like corona. This approach is used in EPOS \cite{Pierog EPOS-LHC}.

The strangeness and baryon production mechanisms predicted by these models are tested through angular correlations between the $\Xi$ baryon (quark content dss) and other hadrons, and in particular those with opposite charge or baryon number. In the string models, strangeness is mostly formed at hadronisation, which results in short-ranged $\Xi$-strangeness correlations, as opposed to the core-corona model where strange quarks are largely formed in the core and can diffuse prior to hadronisation.

\section{Method}
\label{sec:method}
The quantity measured here is the per-trigger yield,
\[
\mathbb{Y}(\Delta y,\Delta\varphi)=\dfrac{1}{N_{\rm trig}}\dfrac{\mathrm{d}^2N_{\rm pairs}}{\mathrm{d}\Delta y\mathrm{d}\Delta\varphi},
\label{eq:per-trigger_yields}
\]
which is the distribution of particle pairs in relative rapidity ($\Delta y$) and azimuthal angle ($\Delta\varphi$) space, divided by the number of triggers $N_{\rm trig}$. To extract the part that is due to balancing charges, same-charge (or baryon number) correlations were subtracted from those of opposite charge.

This was measured for identified $\Xi-\pi$, $\Xi-\rm K$, $\Xi-\rm p$, $\Xi-\Lambda$, and $\Xi-\Xi$ pairs using approximately $8\cdot 10^8$ minimum-bias pp events at $\sqrt{s}=13$ TeV recorded by the ALICE detector \cite{ALICE detector}. Pions, kaons, and protons were identified via the specific energy loss in the Time Projection Chamber and the velocity in the Time-Of-Flight detectors (pseudorapidity $|\eta|<0.8$). The $\Xi$ and $\Lambda$ baryons were reconstructed from their decay products (e.g. $\rm\Xi^-\rightarrow\pi^- + \Lambda$, $\Lambda\rightarrow \pi^-+\rm p$), by making use of their invariant masses and various topological cuts (similar to what is done in Ref. \cite{ALICE strangeness production}). Finally, the multiplicity dependence of the correlation function was measured by dividing the events into multiplicity classes (where the lowest percentiles correspond to the highest multiplicities) based on the response from the V0 counters in the forward regions ($-3.7<\eta<-1.7$ and $2.8<\eta<5.1$).

\section{Results}
\label{sec:results}
\begin{figure}[h!]
\begin{subfigure}{0.499\textwidth}
\centering
\includegraphics[width=0.83\textwidth]{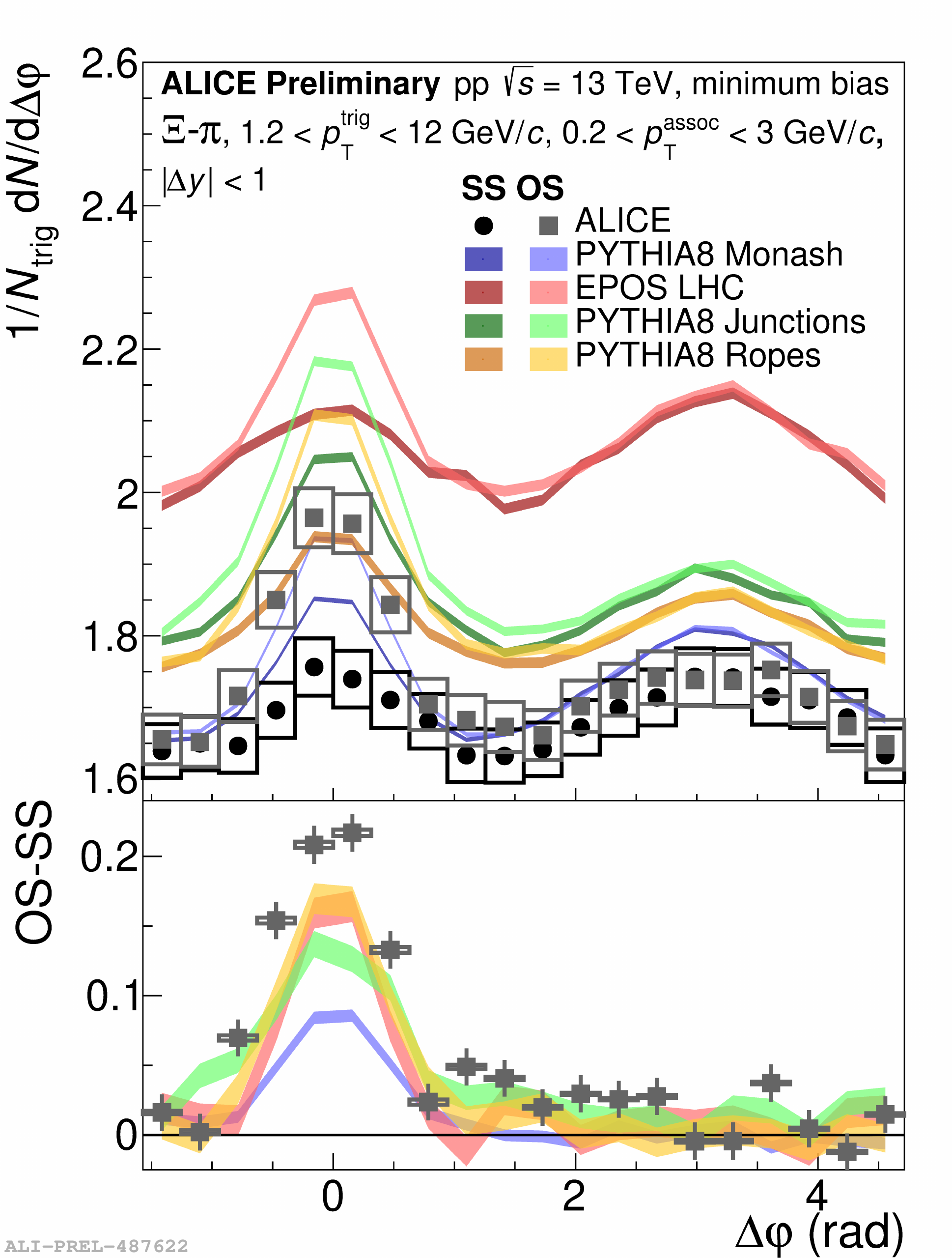}
\end{subfigure}
\begin{subfigure}{0.499\textwidth}
\centering
\includegraphics[width=0.83\textwidth]{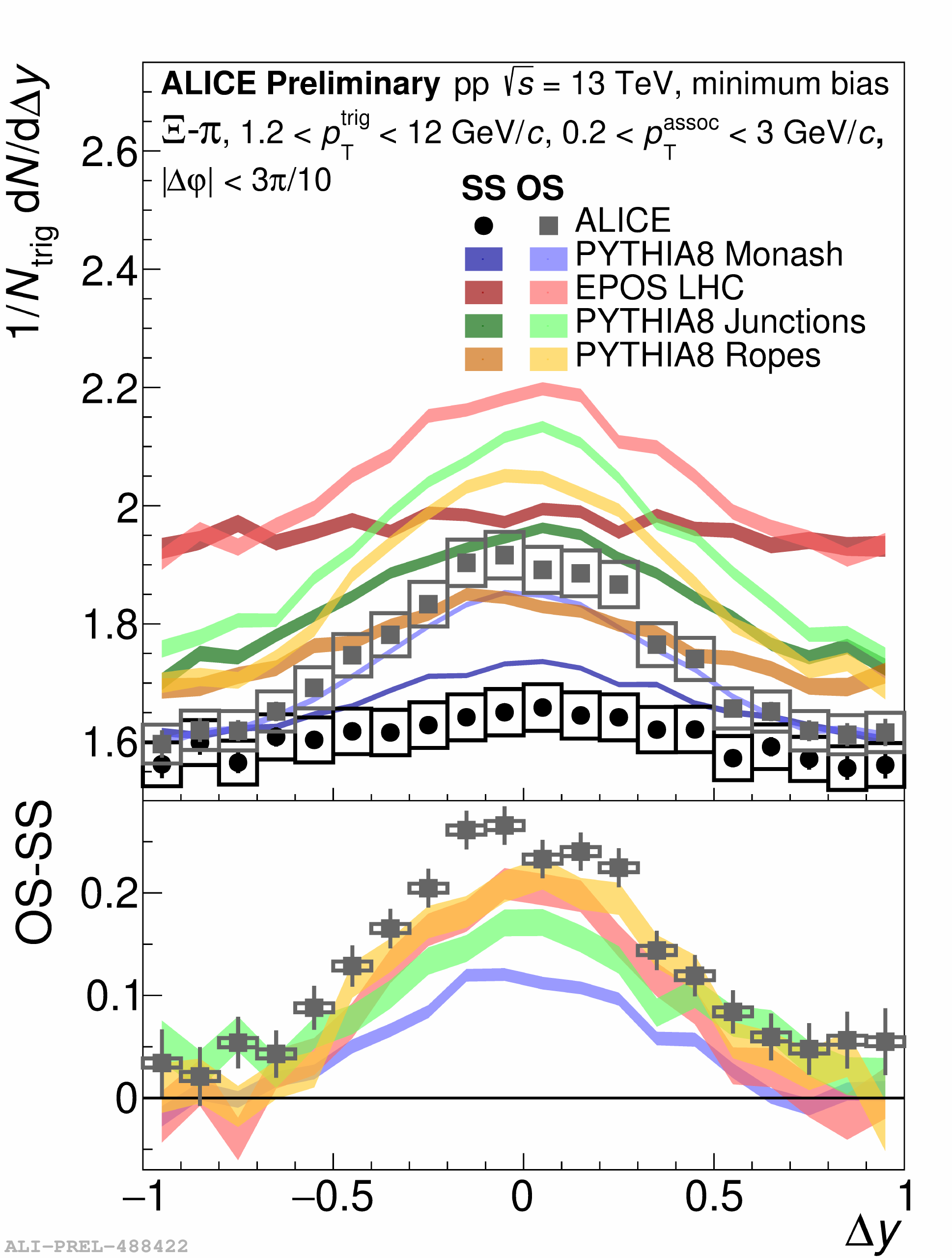}
\end{subfigure}
\caption{Per-trigger yields of $\Xi-\pi$ correlations compared with predictions from PYTHIA8 (including two extensions) and EPOS-LHC, projected onto \textbf{(left)} $\Delta\varphi$ and \textbf{(right)} $\Delta y$ on the near side. The lower panels show the difference between opposite- and same-sign correlations.}
\label{fig:Xi-pi correlations}
\end{figure}

Figure~\ref{fig:Xi-pi correlations} shows angular $\Xi-\pi$ correlations projected on $\Delta\varphi$ and $\Delta y$ (for the near side, $|\Delta\varphi|<$ $3\pi/10$), for both ALICE data and models.
\begin{figure}[h!]
\centering
\includegraphics[width=0.77\textwidth]{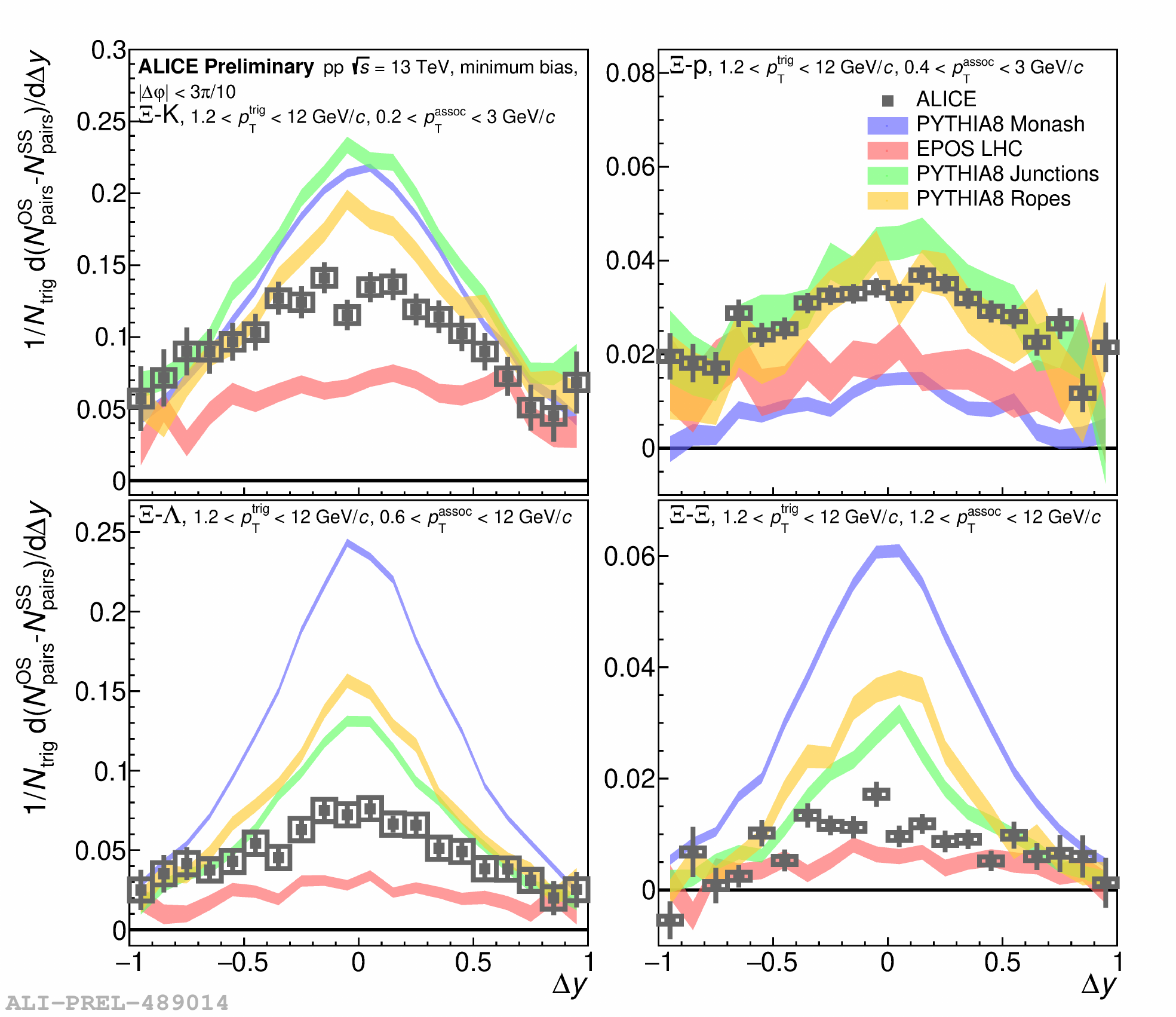}
\caption{Differences between opposite- and same-sign (or baryon number) correlations projected onto $\Delta y$ on the near side, for \textbf{(top left)} $\Xi-\rm K$ pairs, \textbf{(top right)} $\Xi-\rm p$ pairs, \textbf{(bottom left)} $\Xi-\Lambda$ pairs, and \textbf{(bottom right)} $\Xi-\Xi$ pairs.}
\label{fig:corr_diff summary}
\end{figure}
Here, the same-sign correlations are reasonably well described by PYTHIA8 Monash, which can be attributed to good tuning of the single-particle yields, whereas there is a slight offset for the other models. Similar observations have also been made for $\Xi-\rm K$ and $\Xi-\rm p$ correlations. On the other hand, PYTHIA8 Monash underestimates the difference between opposite- and same-sign (OS-SS) correlations, which is better described by the other models or tunes. This means that these models more accurately predict the fraction of the charge from the $\Xi$ baryons that is balanced by pions.

For the other minimum-bias results, only the near-side projections of the OS-SS difference are shown here, which are summarised in Fig.~\ref{fig:corr_diff summary}. Here the predictions from the different models differ significantly, with EPOS-LHC predicting a nearly flat behaviour, which is not observed in data. The reason is that local conservation of quantum numbers is not implemented in EPOS, and this is not necessarily a feature of core-corona models in general. The PYTHIA models, on the other hand, predict a stronger and narrower peak than what is observed in data, except for $\Xi-\rm p$ correlations, where no direct correlations are expected from the diquark breaking mechanism. The junction and rope extensions are better at predicting $\Xi-$baryon correlations, but fail at describing $\Xi-\rm K$ correlations. This favours the additional baryon production mechanism from the junction model, but more development is needed to accurately describe all observations.

\begin{figure}[h!]
\begin{subfigure}{0.325\textwidth}
\centering
\includegraphics[width=\textwidth]{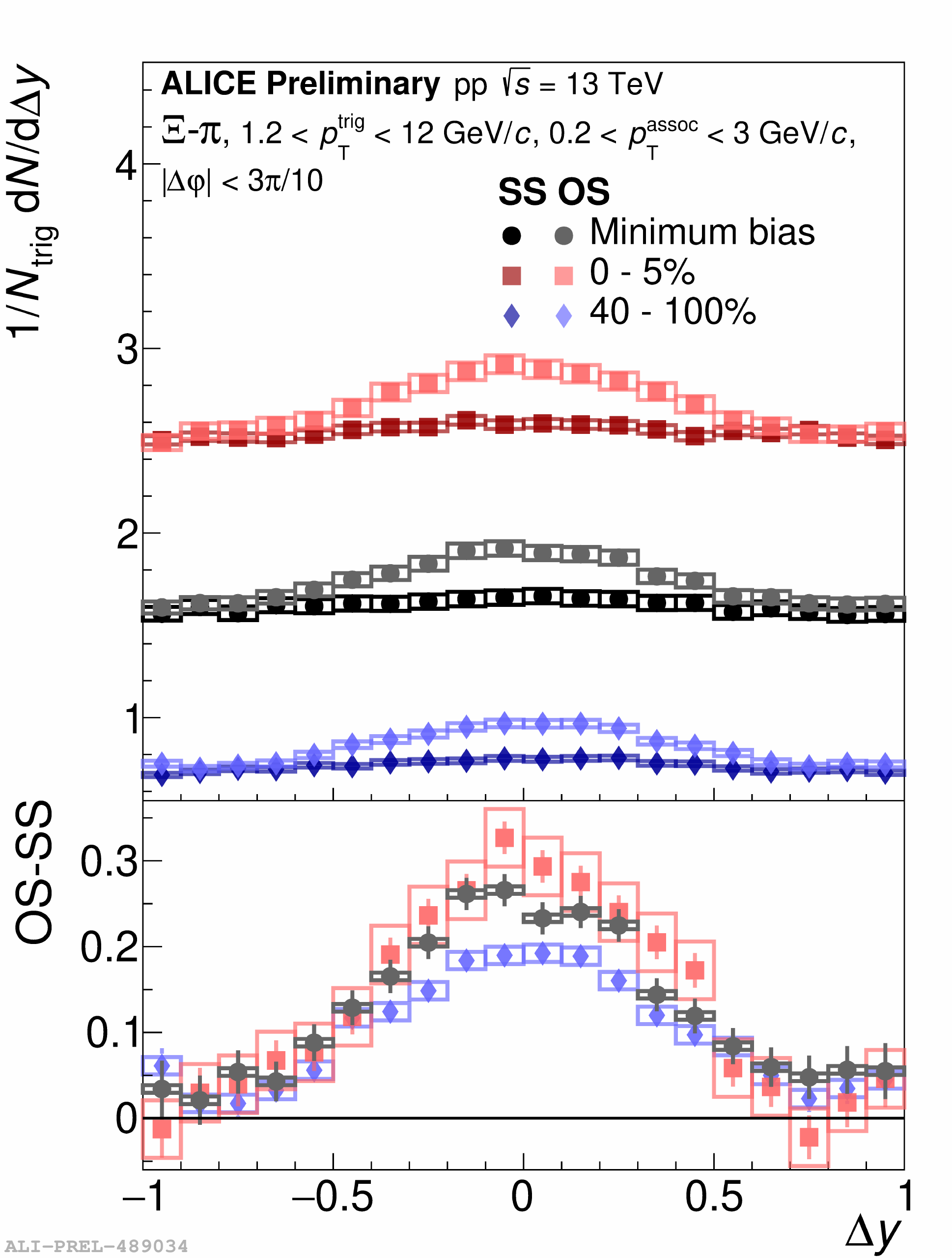}
\end{subfigure}
\begin{subfigure}{0.325\textwidth}
\centering
\includegraphics[width=\textwidth]{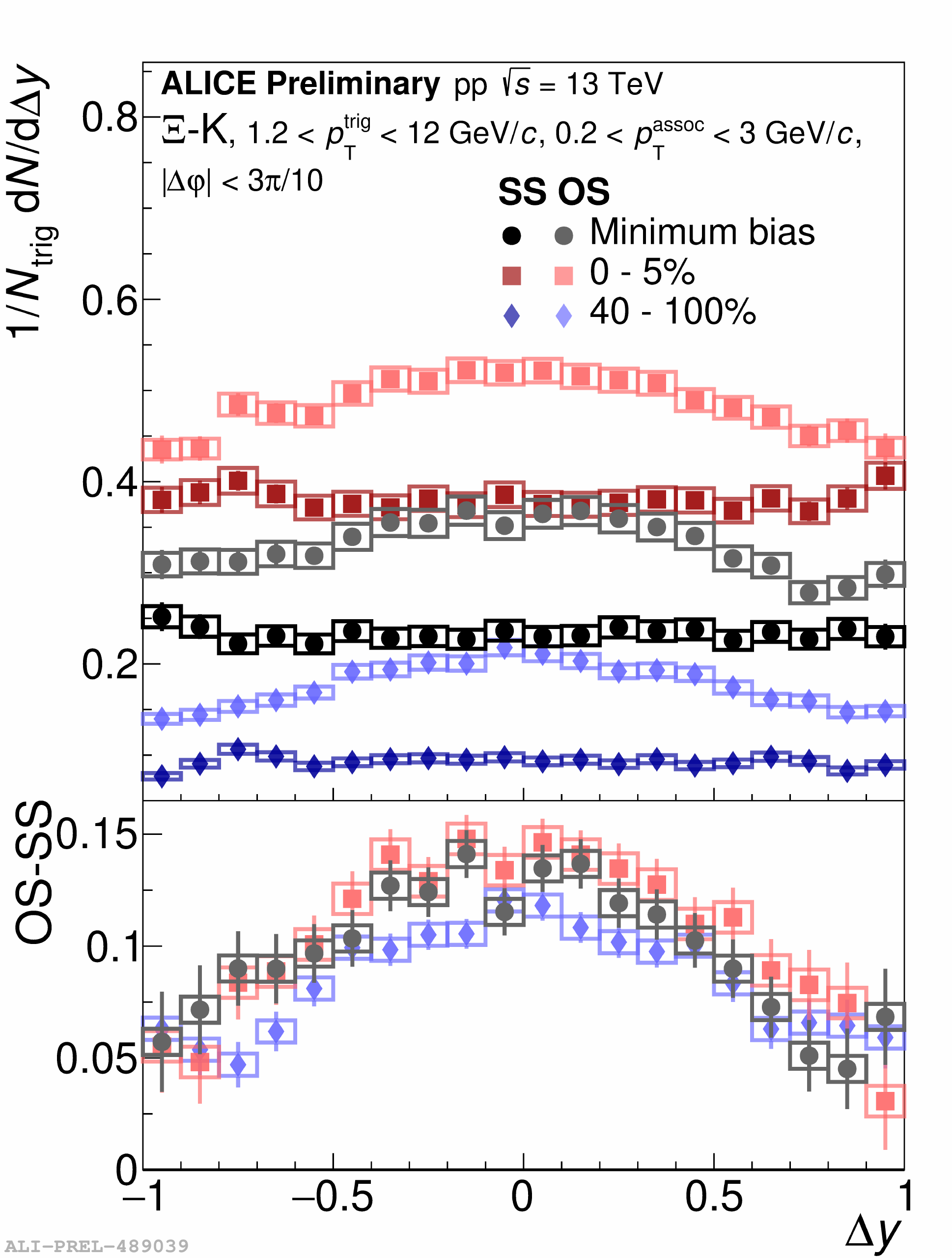}
\end{subfigure}
\begin{subfigure}{0.325\textwidth}
\centering
\includegraphics[width=\textwidth]{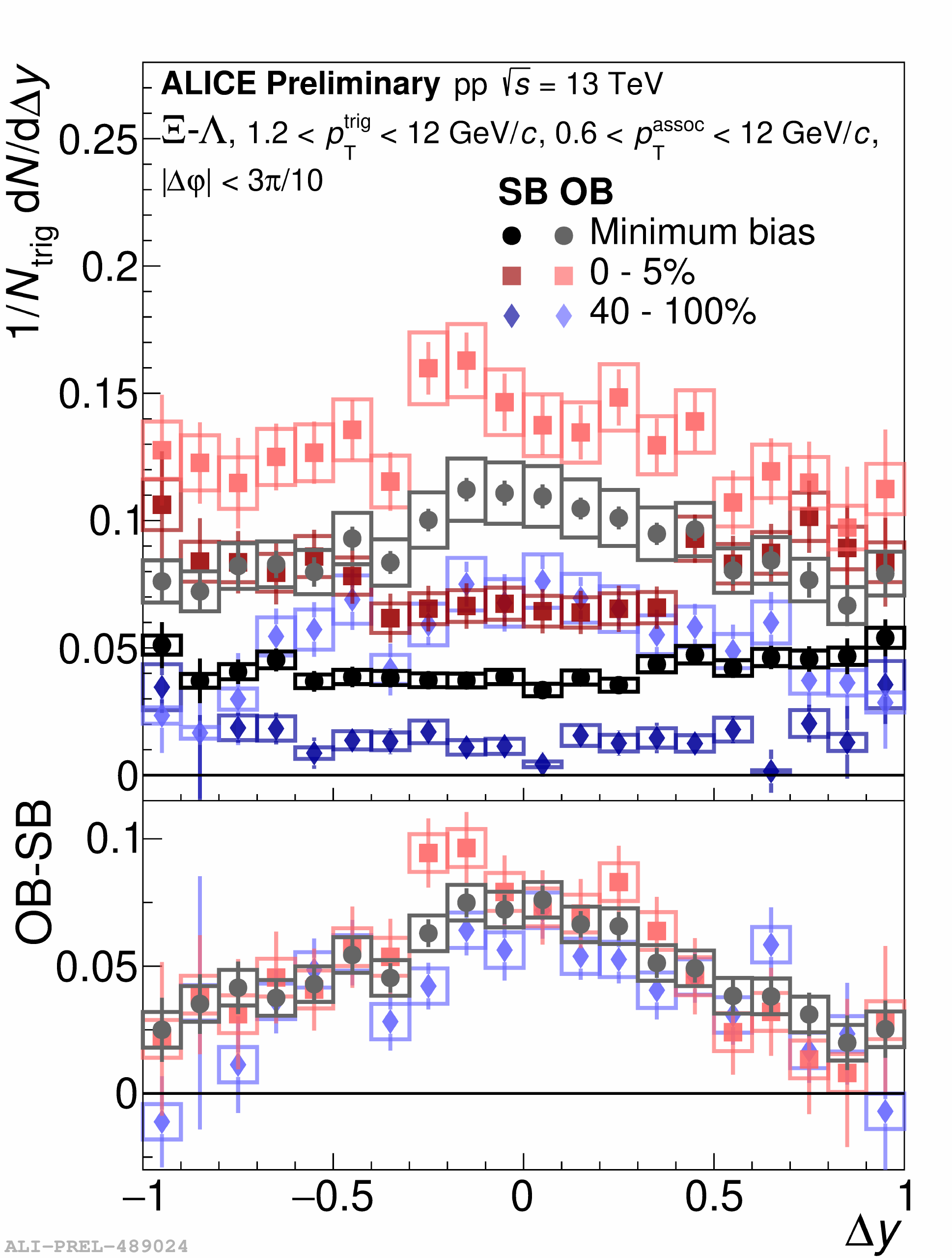}
\end{subfigure}
\caption{Per-trigger yields of \textbf{(left)} $\Xi-\pi$, \textbf{(middle)} $\Xi-\rm K$, and \textbf{(right)} $\Xi-\Lambda$ correlations, measured for three different multiplicity classes and projected onto $\Delta y$ on the near side. The lower panels show the differences between opposite- and same-sign correlations.}
\label{fig:multiplicity dependence}
\end{figure}
Multiplicity dependent $\Xi-\pi$, $\Xi-\rm K$, and $\Xi-\Lambda$ correlations are shown in Fig.~\ref{fig:multiplicity dependence} (similar trends are seen also in $\Xi-\rm p$ and $\Xi-\Xi$ correlations). The shift observed in the baseline with increasing multiplicity is simply due to the increased particle yields, but a small enhancement is also observed for the OS-SS difference, along with some narrowing. A similar behaviour is predicted by PYTHIA, where this can be attributed to radial flow from colour reconnection, so this effect is likely present also in data (although possibly from a different origin). However, the relative difference between multiplicity classes for $\Xi-\Lambda$ correlations is weaker in data than in PYTHIA (not shown), pointing towards a competing mechanism. This could mean that the strangeness production mechanism changes with multiplicity, but more quantitative comparisons with models are needed to conclude in which way.

\section{Conclusions}
\label{sec:conclusions}
While the same-sign $\Xi-$hadron yields predicted by PYTHIA8 Monash largely describe what is observed in data, which is likely due to good tuning of the single-particle yields, the same cannot be said about the OS-SS difference. For this observable, EPOS-LHC predicts an almost flat behaviour, because local conservation of quantum numbers is not implemented in this model, and thus it is difficult to conclude anything about the accuracy of the core-corona approach. On the other hand, the PYTHIA models predict a too strong and narrow OS-SS peak for $\Xi-$strangeness correlations, which means that the strange quarks seem to have diffused somewhat in data prior to hadronisation. Finally, for $\Xi-$baryon correlations the junction extension describes the data better than the Monash tune, favouring this mechanism for producing baryons.

The multiplicity dependence seems to be dominated by radial flow, but some observations indicate a competing mechanism, possibly related to changes in the $\Xi$ production. However, more work is required to conclude how.

\end{document}